\documentclass[showpacs,aps,pre,twocolumn]{revtex4}
\usepackage{graphicx}
\usepackage{amsfonts}

\begin{document}
\title{Transmission of matter wave solitons through nonlinear traps and barriers}
\author{Josselin Garnier\dag\footnote[7]{Corresponding author (garnier@math.jussieu.fr)} \, and  Fatkhulla Kh. Abdullaev\ddag }
\affiliation{\dag\ Laboratoire de Probabilit\'es et Mod\`eles
Al\'eatoires \& Laboratoire Jacques-Louis Lions, Universit{\'e}
Paris VII, 2 Place Jussieu, 75251 Paris Cedex 5, France\\
\ddag\ Physical-Technical Institute of the Uzbekistan Academy of
Sciences, 700084, Tashkent-84, G.Mavlyanov str.,2-b, Uzbekistan}

\begin{abstract}
The transmissions of matter wave solitons through linear and
nonlinear inhomogeneities induced by the spatial variations of the
trap and the  scattering length in Bose-Einstein condensates are
investigated. New  phenomena,  such as the enhanced transmission
of a soliton through a linear trap  by a modulation of the
scattering length, are exhibited. The theory is based on the
perturbed Inverse Scattering Transform for solitons, and we show
that radiation effects are important. Numerical simulations of the
Gross-Pitaevskii equation confirm the theoretical predictions.
\end{abstract}
\maketitle

\newpage

\section{Introduction}
The transmission of matter wavepackets through  inhomogeneities of
different types of Bose-Einstein condensates (BECs) has recently
attracted a lot of attention, because this phenomenon is important
for the design of control methods of the soliton parameters and
atomic soliton lasers \cite{Garcia}. The transmission and
reflection of bright and dark matter wave solitons has been
studied in the case of linear inhomogeneities, induced by the
variations in space of the potential field
\cite{KM,FrantzKiv,Brand,GHW}. The effect of a potential step or
impurity, including the soliton train evolution, has been analyzed
in \cite{Seaman}. The case of inhomogeneities produced by spatial
variations of the scattering length  has been  less investigated
\cite{AS,AGKT,Fibich}. Numerical simulations of the problem of
soliton propagation have been carried out, when the linear and
nonlinear inhomogeneities compete with each other. It was found
that for some choices of parameters the enhanced transmission of
soliton through the nonlinear barrier is possible \cite{kevrekidis2}.

The purpose of this work is to develop the theory describing the
transmission of matter wave solitons through nonlinear barriers.
Such barriers can be produced by using the Feshbach resonance
method, namely  by the local variation of the external magnetic
field $B(x)$ in space near the resonant value $B_c$
\cite{Pitaevskii}. By the small variation of the field near the
resonant value we can induce the large variations of the
scattering length in space according to the formula
$$
a_{s}(x) = a_{b}\left( 1 - \frac{\Delta}{B_c - B(x)}\right),
$$
where $a_b$ is the background value of the atomic scattering
length and $\Delta$ is the resonance width.
Optical methods for manipulating the value of the
scattering length are also possible \cite{FKSW}. As a result, the  mean
field nonlinear coefficient (which is proportional to the
scattering length $a_s$) in the Gross-Pitaevskii equation has
a spatial dependence. We will use the perturbed Inverse Scattering
Transform theory (see for
example \cite{Karpman,AbdullaevB,Garnier98}) to describe the
transmission of bright matter wave solitons through the nonlinear
barriers. This approach allows us to analyze the adiabatic
dynamics of solitons as well as the radiative processes during the
soliton propagation through inhomogeneities.

\section{The model}
The quasi one-dimensional  Gross-Pitaevskii (GP) equation describing the
wavefunction of  BEC  in an elongated trap has the form \cite{perez}
\begin{equation}
i\hbar\psi_t + \frac{\hbar^2}{2m}\psi_{xx} - V(x) \psi -
 g_{1D}(x)|\psi|^2 \psi = 0.
\end{equation}
Here $g_{1D}(x)= 2\hbar\omega_{\perp}a_{s}(x)$, where
$\omega_{\perp}$ is the transverse oscillator frequency,
$a_{s}(x)$ is the spatially dependent atomic scattering length,
and $V(x)$ is the linear potential. Both $a_s(x)$ and $V(x)$ are
assumed to be constant outside a given domain, where we assume
that the  scattering length takes the constant negative value
$a_{s0}$ and $V$ is zero, $\int |\psi|^2 dx = N$, where $N$ is the
number of atoms. We denote by $\bar{g}_{1D}=
2\hbar\omega_{\perp}a_{s0}$ the reference value of the nonlinear
coefficient.

From  now on we express $x$ in units of the healing length $\xi = \hbar / \sqrt{ n_0 \bar{g}_{1D}  m}$ and $t$ in units of $t_{0}= \xi/(2c)$,
where $c = n_0 \bar{g}_{1D} / m$ is the Bogoliubov speed of sound
and $n_0$ the peak density.
We also normalize the mean field wavefunction by $\sqrt{n_0}$,
so that we obtain the dimensionless
GP equation in the form of the Nonlinear Schr\"odinger (NLS) equation
\begin{equation}
\label{eq:cq} i u_t + u_{xx} + \gamma(x) |u|^2 u  - V_l(x) u= 0 \, ,
\end{equation}
with $\gamma(x) = 2 -  V_{nl} (x) = 2 a_s(x) / a_{s0}$.
If $V_l=V_{nl}=0$, then this equation
can be reduced to the standard NLS equation
that supports soliton solutions.
The bright soliton solution is:
$$
u_s (x,t) = 2\nu\mbox{sech}[2\nu(x - x_s(t))]
\exp[ 2
i\mu (x -x_s(t) )  + i \phi_s(t) ].
$$
The soliton amplitude is $2\nu$ and its velocity is $4 \mu$.
The soliton center and phase $x_s(t)$ and $\phi_s(t)$
satisfy
$$
\frac{dx_s}{dt} = 4 \mu , \ \ \ \ \ \frac{d \phi_s}{dt} = 4 (\nu^2
+ \mu^2).
$$
The matter wave soliton moving in the linear and nonlinear
potentials $V_l$ and $V_{nl}$ experiences velocity and mass
modulations and emits radiation. To describe this process we use
the perturbation theory based on the Inverse Scattering
Transform (IST).

\section{Quasi-particle approach}
\label{sec:quasipart}%
Applying the first-order perturbed IST theory, we obtain the
system of equations for the soliton amplitude and velocity
$$
\frac{d \nu}{dt}=0, \ \ \ \ \ \ \frac{d\mu}{dt} = -\frac{1}{4 \nu}
W'(\nu,x_s), \ \ \ \ \ \ \frac{d x_s}{dt} = 4 \mu ,
$$
where the prime stands for a derivative with respect to $x$ and the
effective potential  has the form
$$
W(\nu,x) = W_l(\nu,x)+W_{nl}(\nu,x),
$$
with
\begin{eqnarray}
\label{def:wl}
&&
W_l(\nu,x) =  \nu \int_{-\infty}^{\infty} \frac{1}{\cosh^2(z)} V_l(\frac{z}{2\nu}+x) dz, \\
\label{def:wnl} && W_{nl}(\nu,x) = 2 \nu^3 \int_{-\infty}^{\infty}
\frac{1}{\cosh^4(z)} V_{nl}(\frac{z}{2\nu}+x) dz.
\end{eqnarray}
We can thus write the effective equation describing the dynamics of the soliton
center as a quasi-particle moving in the effective potential $W$:
\begin{equation}
\nu_0 \frac{d^2 x_s}{dt^2} = - W'(\nu_0 ,x_s),
\end{equation}
where $4\nu_0$ is the mass (number of atoms) of the incoming
soliton. This system has the integral of motion
\begin{equation}
\label{ener0} \frac{\nu_0}{2} \left(\frac{dx_s}{dt}\right)^2 +
W(\nu_0,  x_s) = 8 \nu_0 \mu_0^2,
\end{equation}
where $4\mu_0$ is the velocity of the incoming soliton.
Note that this approach is standard and it was recently applied
 in \cite{kevrekidis}.
In this work, it is the first step as we will include second-order and
radiation effects in the next section.
Let us briefly discuss the main results that can be obtained with
the quasi-particle approach.

{\bf Barrier potential.} Let us first examine the case where the
potential $V$ is a barrier, meaning that $V \geq 0$ and $\lim_{|x|
\rightarrow \infty} V(x)=0$. When the soliton approaches the
barrier, it slows down, and it eventually goes through the barrier
if its input energy is above the maximal energy barrier, meaning
$8 \nu_0 \mu_0^2  > W_{max}(\nu_0)$. For a linear barrier
potential which is an even function, we have $W_{max}(\nu) =
W_l(\nu,0)$, where $W_l$ is given by (\ref{def:wl}). For a
nonlinear barrier potential which is an even function, we have
$W_{max}(\nu)= W_{nl}(\nu,0)$. After passing through the barrier,
the soliton recovers its initial mass and velocity. In that sense,
the transmission coefficient is one.

If, on the contrary, the velocity of the incoming velocity is such
that $8 \nu_0 \mu_0^2 < W_{max}(\nu_0)$, then the soliton is
reflected by the barrier. After the interaction with the barrier,
the soliton velocity takes the value $-4 \mu_0$.
In that sense, the transmission coefficient is zero.\\

{\bf Trap potential.} We now examine the case where the potential
is a trap, meaning that $V \leq 0$ and $\lim_{|x| \rightarrow
\infty} V(x)=0$. When the soliton approaches the trap, it speeds
up, and it eventually goes through the barrier whatever its
initial velocity is. This is the prediction of the quasi-particle
approach. However, we shall see that the interaction with the trap
generates radiation and reduces the mass and energy of the
soliton. As a result, the soliton may not be able to escape the
trap if its initial velocity is too small. We shall discuss this
point  in the next section.

\section{Radiation effects}
The properties of the radiation emitted by the soliton interacting with 
the inhomogeneities are defined by the Jost
coefficients $a(t,\lambda)$ and $b(t,\lambda)$ of the associated
linear spectral problem for the NLS equation
\cite{Karpman,AbdullaevB,Garnier98}. We assume that the linear and
nonlinear potentials are localized functions, and we denote by
$\hat{V}(k) = \int V(x) \exp(i k x) dx$ their Fourier transforms.
If the soliton goes through the potential, then we find
\begin{widetext}
$$
\frac{b}{a} (t=+\infty ,\lambda) = \frac{-i\pi (\lambda -\mu-i \nu)^2}{16 \mu^3 \cosh
\left( \frac{\pi}{4} \frac{\lambda^2 +\nu^2-\mu^2}{\mu \nu} \right)}
\times \left[ \hat{V}_{l} \left( k(\lambda,\nu,\mu) \right)+
\frac{[(\lambda+\mu)^2 + \nu^2][ (\lambda-3\mu)^2+8 \mu^2+\nu^2]}{12\mu^2}
\hat{V}_{nl} \left( k(\lambda,\nu,\mu) \right)\right]
$$
\end{widetext}
where
$$
k(\lambda, \nu,\mu) = \frac{ (\lambda-\mu)^2 + \nu^2}{\mu}
$$
The total radiated mass density during the interaction with the potentials is
$$
n( \lambda) \simeq \frac{1}{\pi}  \left|  \frac{b}{a} (t=+\infty ,\lambda) \right|^2
$$
The total mass (number of atoms) and Hamiltonian are preserved:
\begin{eqnarray*}
&&{\cal N}= \int_{-\infty}^{\infty} |u|^2 dx  \\
&&{\cal H} = \int_{-\infty}^{\infty} [|u_x|^2 -  |u|^4 +V_l(x)
|u|^2 +\frac{1}{2} V_{nl}(x) |u|^4 ] dx.
\end{eqnarray*}
They can be expressed in terms of the radiation and soliton components
as
\begin{eqnarray*}
&&{\cal N} = 4 \nu + \int_{-\infty}^{\infty} n(\lambda)  d \lambda , \\
&&{\cal H} = 16 \nu \mu^2 -\frac{16}{3} \nu^3 + 2 W(\nu,x_s) + 4
\int_{-\infty}^{\infty} \lambda^2 n(\lambda) d \lambda .
\end{eqnarray*}
Note that this expression of the Hamiltonian
 generalizes the expression of the energy (\ref{ener0}).
The coefficients $(\nu_T,\mu_T)$ of the transmitted soliton are
\begin{eqnarray}
\label{nu1}
&&\nu_T = \nu_0 - \frac{1}{4} \int_{-\infty}^{\infty} n(\lambda)d\lambda, \\
\label{mu1} &&\mu_T = \mu_0  - \frac{1}{8} \int_{-\infty}^{\infty}
\left( \frac{\lambda^2}{\mu_0 \nu_0}
+\frac{\nu_0}{\mu_0}-\frac{\mu_0}{\nu_0} \right)
n(\lambda)d\lambda .
\end{eqnarray}
These formulas allow us to study
and characterize the transmission
of a soliton through a general barrier in various regimes.
We can consider the transmission/reflection
of a bright soliton through a nonlinear
barrier, the transmission/trapping in a nonlinear trap,
and competition effects for the transmission
through the superposition of  nonlinear and  linear potentials.

In the next subsections, we compare our theoretical predictions
with results from numerical simulations of the one-dimensional GP
equation, with a Gaussian linear potential and/or a nonlinear
Gaussian potential:
$$
V_l (x) = V_{l0} \exp\left( - \frac{x^2}{x_c^2}\right),
\hspace*{0.2in} V_{nl} (x) = V_{nl0} \exp\left( -
\frac{x^2}{x_c^2}\right).
$$
We consider the transmission
of a soliton incoming from the left homogeneous
half-space with parameters $(\nu_0,\mu_0)$.
We plot the variations of the soliton  parameters versus
the value of the velocity for $\nu_0=0.5$ and $x_c=0.5$.
We consider different combinations of linear and nonlinear potentials.\\

{\bf Nonlinear barrier.}
We consider the case $V_{l0}=0$ and $V_{nl0}>0$.
We have seen in Section \ref{sec:quasipart}
that the condition
for the transmission through a nonlinear barrier is that
the velocity of the incoming velocity is large enough so that
$8 \nu_0 \mu_0^2 > W_{max}(\nu_0)$.
Note that this means that the critical velocity parameter $\mu_{crit}$,
defined by the identity $8 \nu_0 \mu_{crit}^2 =W_{max}(\nu_0)$,
is of the order of $V_{nl0}^{1/2}$ for $V_{nl0}\ll 1$.
If the transmission condition fails, then the soliton is reflected.

\begin{figure}
\begin{center}
\begin{tabular}{c}
\includegraphics[width=5.7cm]{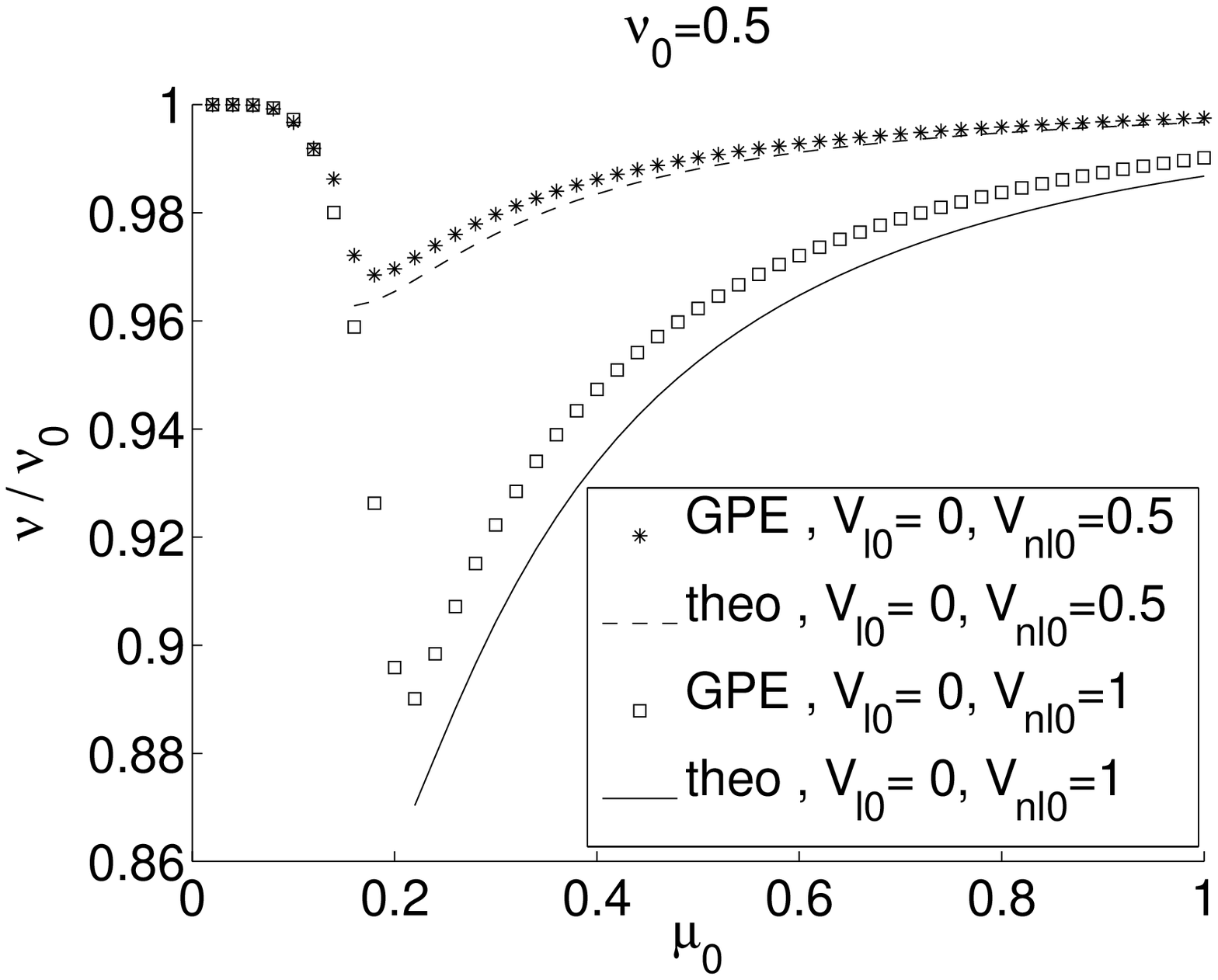}\\
\includegraphics[width=5.7cm]{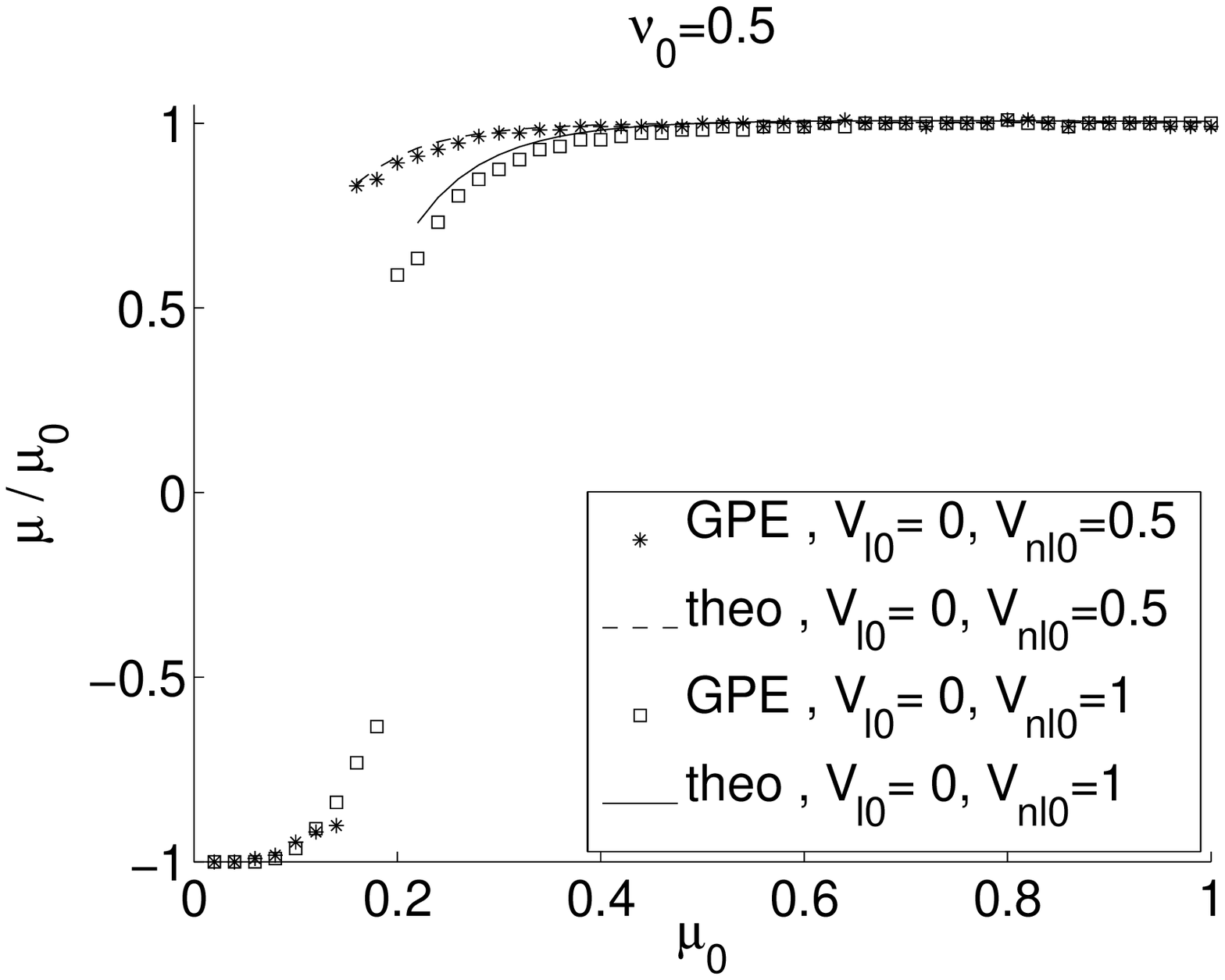}
\end{tabular}
\vspace*{-0.25in}
\end{center}
\caption{Soliton parameters variations versus input velocity parameter $\mu_0$
for a nonlinear barrier. The theoretical reflection condition is
$8 \nu_0 \mu_0^2 < W_{max}(\nu_0)$ which gives $\mu_{crit} = 0.21$ for $V_{nl0}=1$ and
$\mu_{crit}=0.15$ for $V_{nl0}=0.5$,
in excellent agreement with the simulations.
The predicted mass and velocity reductions are also
quantitatively accurate, especially for $V_{nl0}=0.5$.
\label{fig1} }
\end{figure}

These results are obtained in the quasi-particle approach and neglect
the radiation emission phenomemon.
Taking into account radiation yields that the transmission is not complete
in the case $8 \nu_0 \mu_0^2 > W_{max}(\nu_0)$, in the sense that
the transmitted soliton mass is not equal to the
incoming soliton mass (see Figure \ref{fig1}).
The mass loss is described by (\ref{nu1}),
and it is of the order of $V_{nl0}$ for $V_{nl0}\ll 1$.\\

{\bf Nonlinear trap.} We consider the case $V_{l0}=0$ and
$V_{nl0}<0$. The quasi-particle approach predicts full soliton
transmission, but neglects radiation phenomena. When taking into
account radiation emission, we can exhibit the mass and energy
loss during the interaction with the trap potential. The losses
are described by (\ref{nu1}-\ref{mu1}) and are accurate for an
initial velocity large enough. Indeed, if the initial velocity is
not large enough, then the energy loss during the interaction with
the potential does not allow the soliton to escape the trap.

\begin{figure}
\begin{center}
\begin{tabular}{c}
\includegraphics[width=5.7cm]{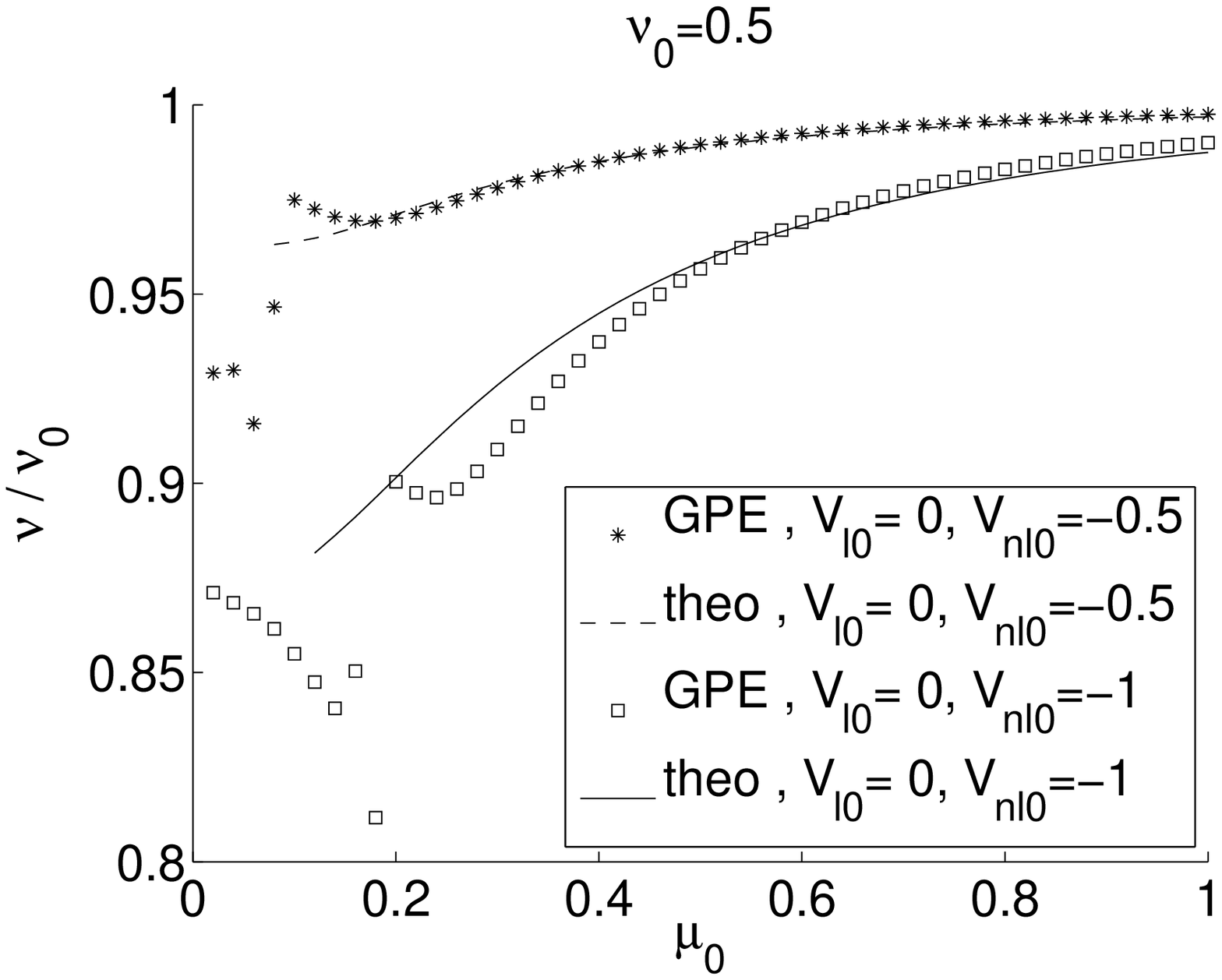}\\
\includegraphics[width=5.7cm]{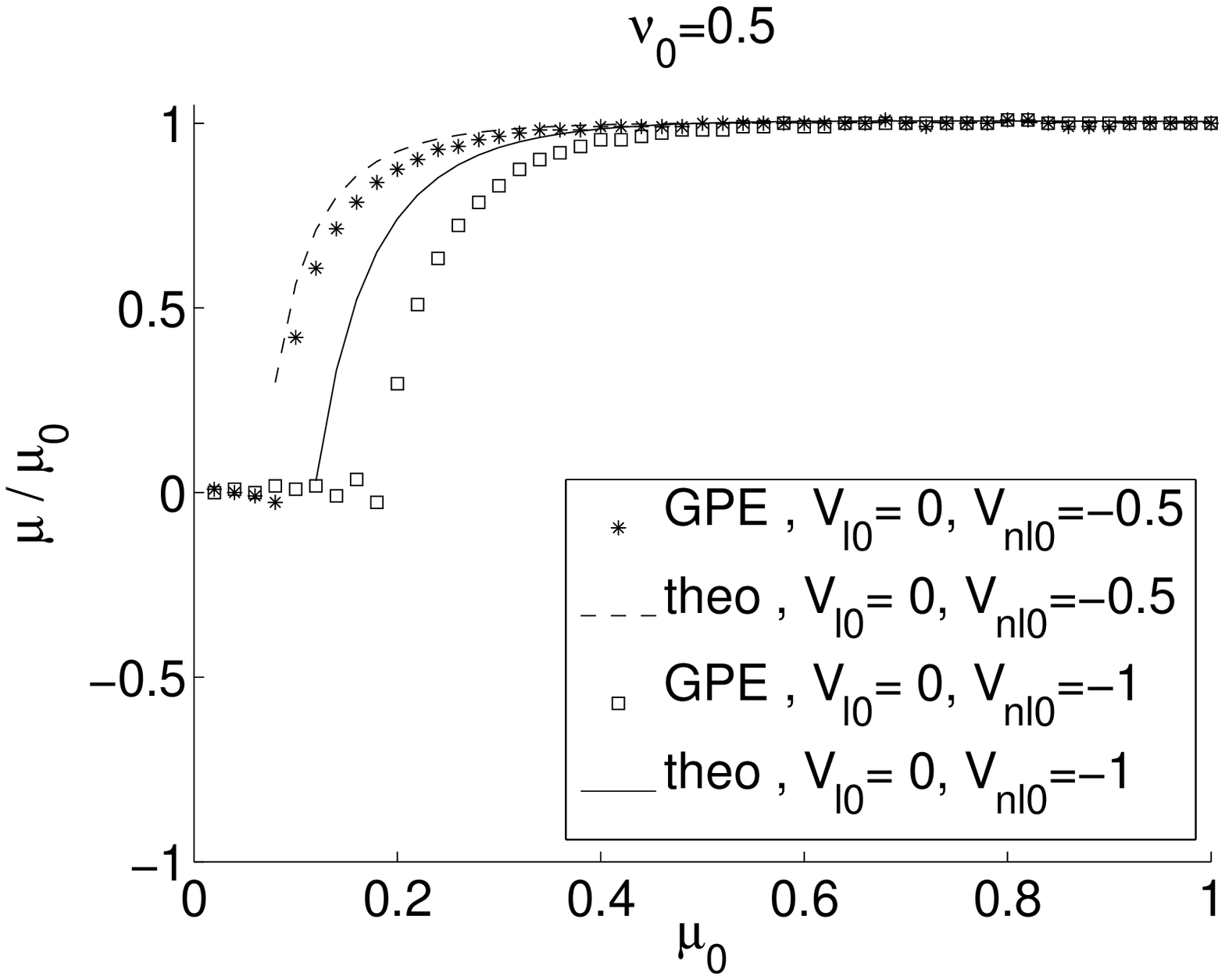}
\end{tabular}
\vspace*{-0.25in}
\end{center}
\caption{Soliton parameters variations versus input velocity parameter $\mu_0$
for a nonlinear trap.
The theoretical trap condition is $\mu_T<0$
which gives $\mu_{crit} = 0.12$ for $V_{nl0}=-1$ and
$\mu_{crit}=0.07$ for $V_{nl0}=-0.5$.
The agreement with the simulations is noticeable for $V_{nl0}=-0.5$,
while the case $V_{nl0}=-1$ is only in qualitative agreement with the simulations.
\label{fig2} }
\end{figure}

For small initial velocity, the expression for the radiation
is not precise enough, because the velocity experiences a
modulation which is relatively large compared to its
initial value.
When the soliton center is $x$,
the soliton velocity is given by
$4 \mu(x)$ with
$$
\mu(x) = \sqrt{\mu_0^2 - \frac{W_{nl}(\nu_0,x)}{16 \nu_0}}.
$$
Its maximal value is $ \mu_{max} =  \sqrt{\mu_0^2 -
{W_{nl}(\nu_0,0)}/({16 \nu_0})} $ where $W_{nl}(\nu_0,x) <0$ is
given by (\ref{def:wnl}). During the interaction with the barrier,
the soliton velocity is about $4 \mu_{max}$, so the energy loss
can be estimated by $-4 \int \lambda^2 n(\lambda) d \lambda$,
where $\mu_{max}$ is substituted for $\mu_0$ in the expression of
$b/a$. If a negative value $\mu_T$ is obtained in (\ref{mu1}),
then this means that the soliton has not been transmitted, but was
trapped by the nonlinear potential (see Figure \ref{fig2}). Note
that $n(\lambda) \sim V_{nl0}^2$ so that  the critical value
$\mu_{crit}$ for trapping is of the order of $V_{nl0}$.

If the initial velocity is large enough to ensure transmission,
then the soliton emits radiation and looses mass.
The mass loss is described by (\ref{nu1})
and it is of the order of $V_{nl0}$ for $V_{nl0}\ll 1$.\\

{\bf Enhanced transmission by nonlinear modulation.}
As pointed out in \cite{kevrekidis},
the nonlinear potential $V_{nl}$ can help a soliton going through
a trap potential $V_l$.
We illustrate this assertion based on numerical experiments in this section
and justify it with our perturbed IST approach.

\begin{figure}
\begin{center}
\begin{tabular}{c}
\includegraphics[width=5.7cm]{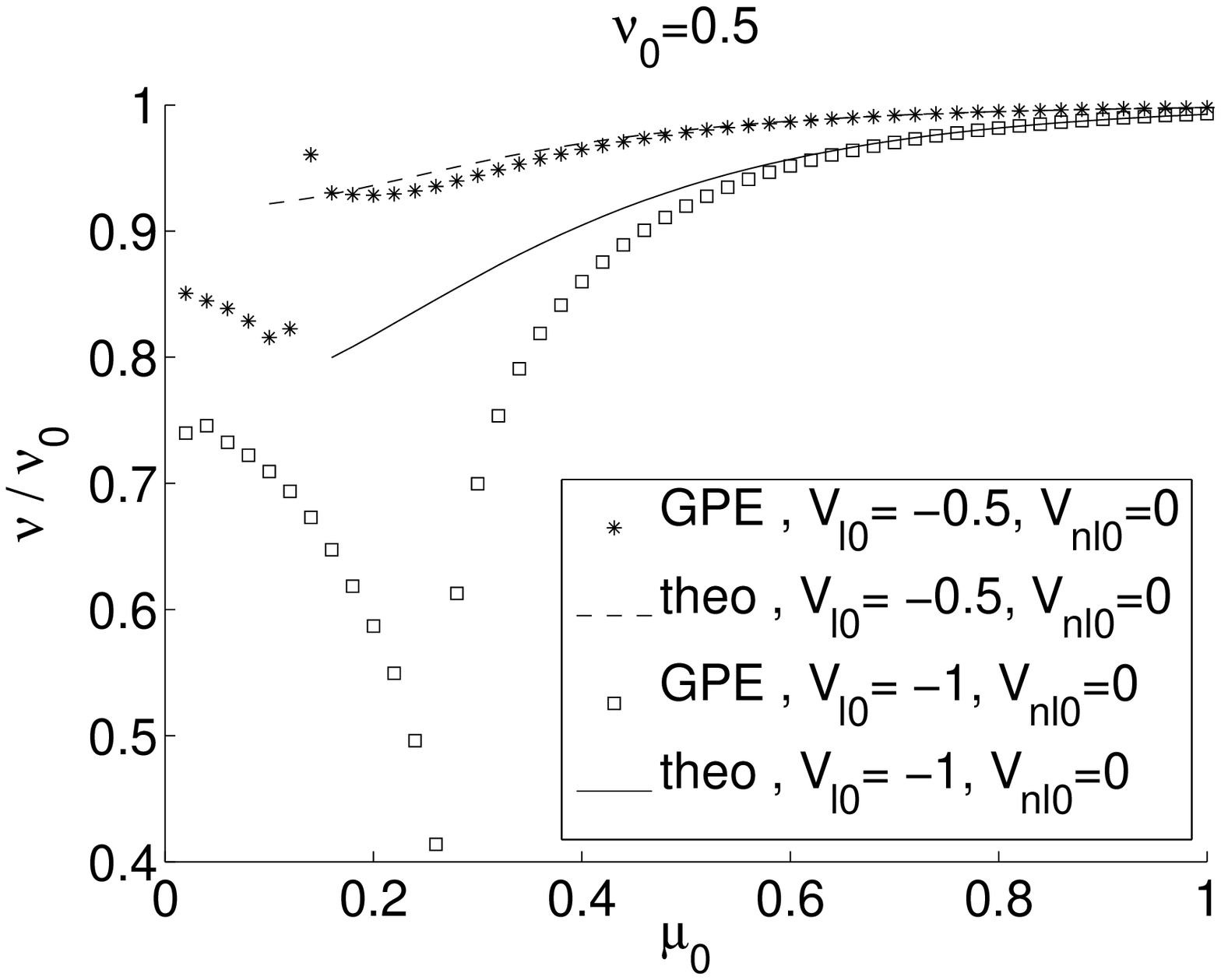}\\
\includegraphics[width=5.7cm]{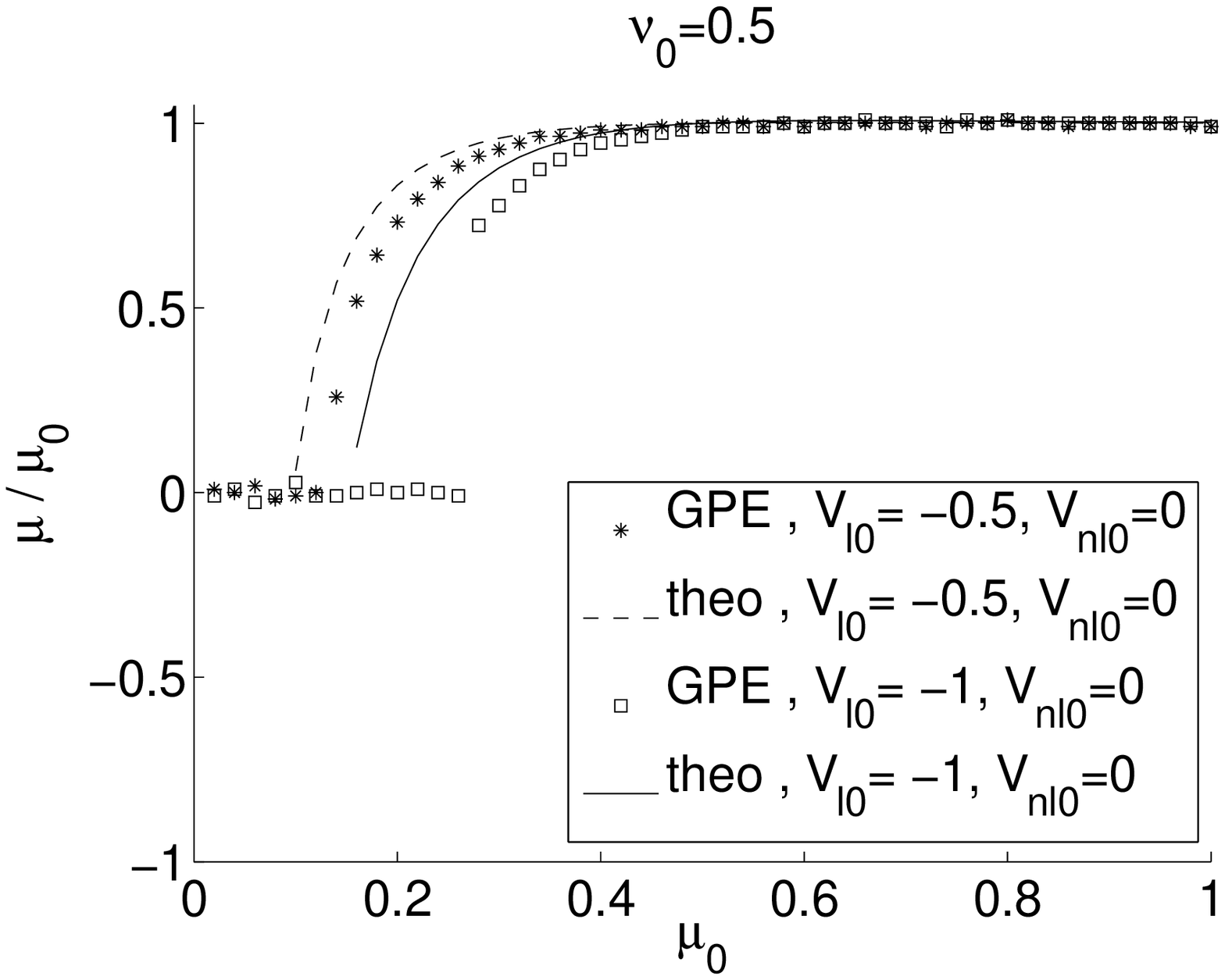}
\end{tabular}
\vspace*{-0.25in}
\end{center}
\caption{Soliton parameters variations versus input velocity parameter $\mu_0$
for a linear trap.
The theoretical trap condition is
$\mu_T<0$
which gives $\mu_{crit} = 0.15$ for $V_{l0}=-1$ and
$\mu_{crit}=0.1$ for $V_{l0}=-0.5$,Ê
in good agreement with the simulations  in the case $V_{l0}=-0.5$.
\label{fig3} }
\end{figure}

\begin{figure}
\begin{center}
\begin{tabular}{c}
\includegraphics[width=5.7cm]{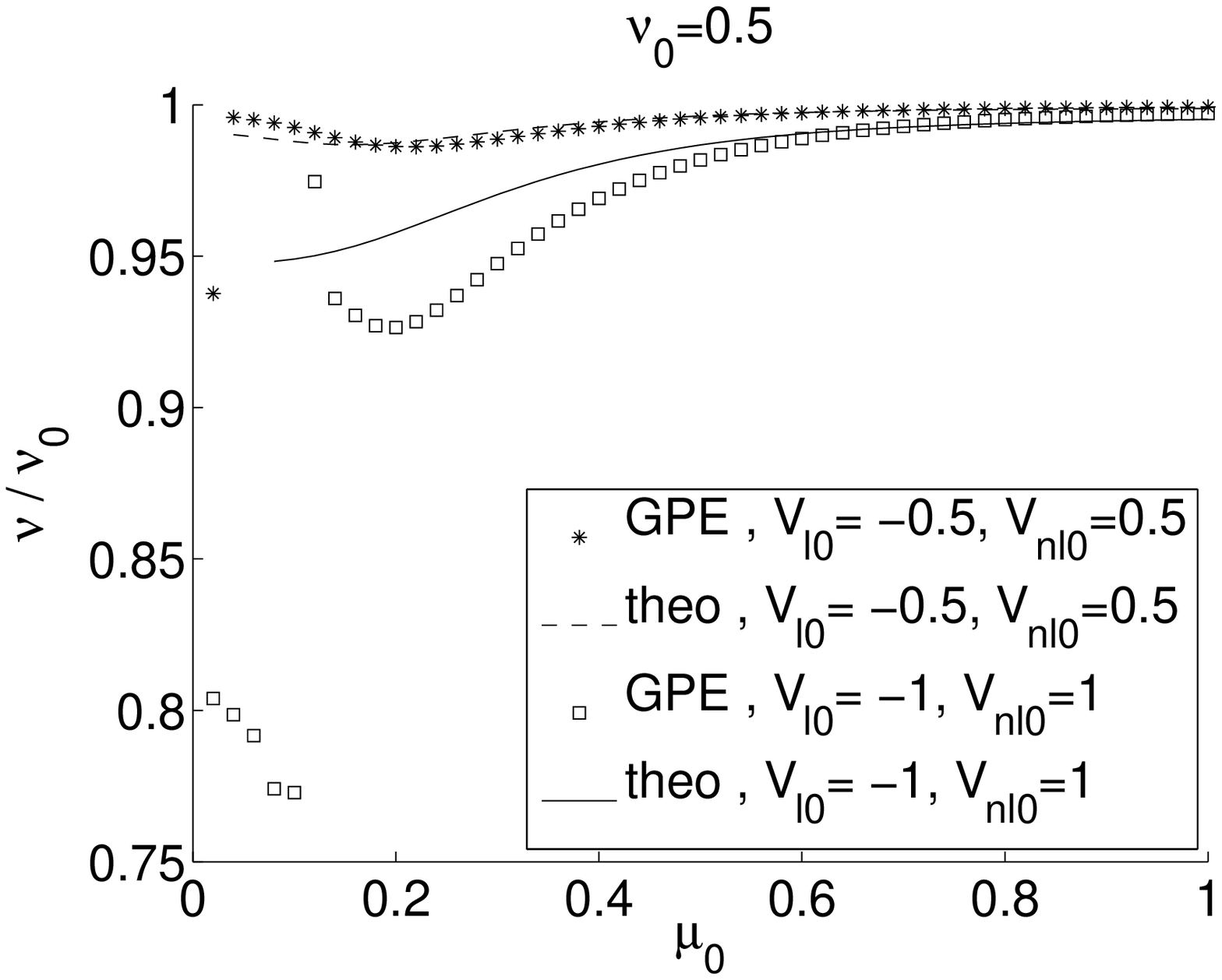}\\
\includegraphics[width=5.7cm]{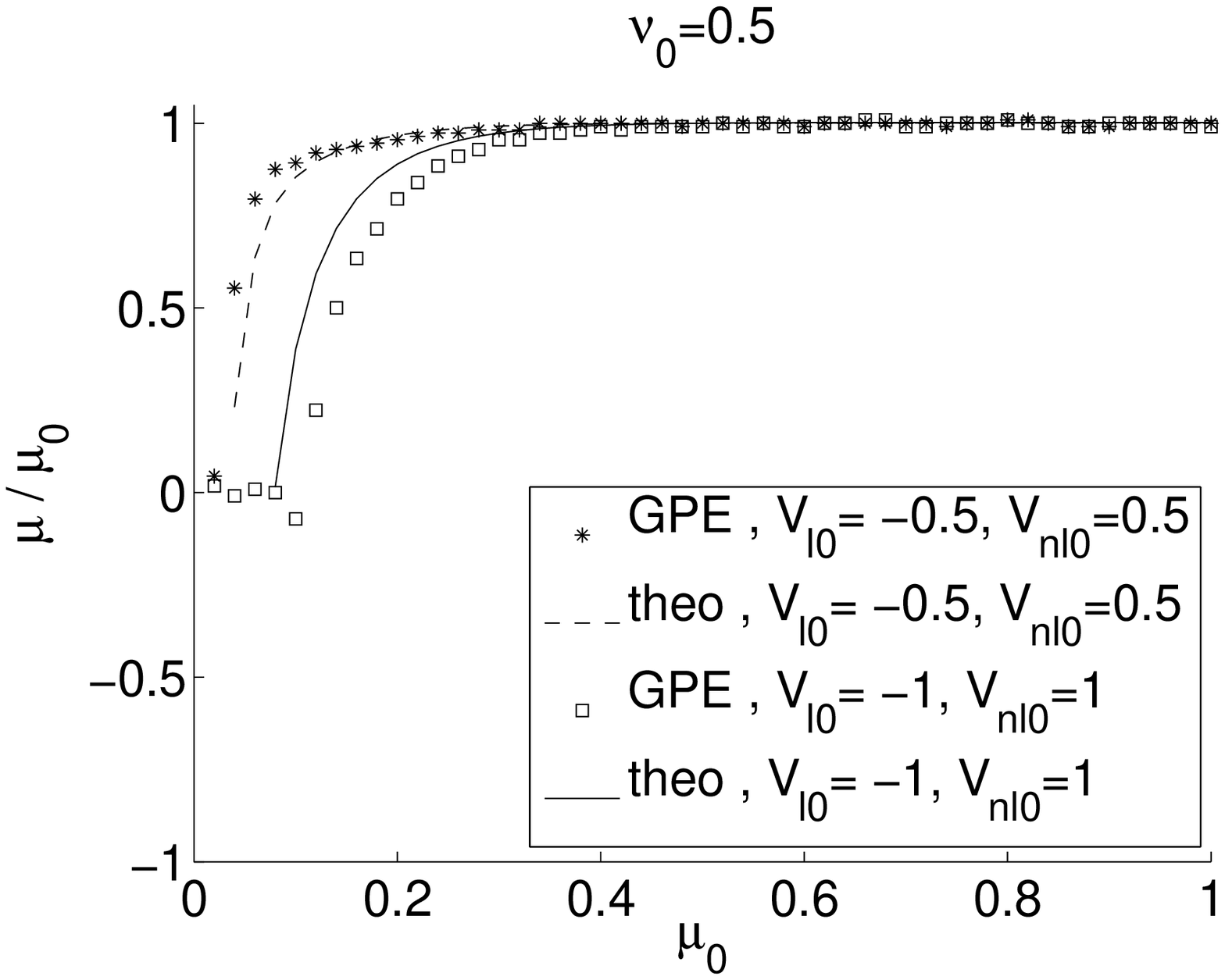}
\end{tabular}
\vspace*{-0.25in}
\end{center}
\caption{Soliton parameters variations versus input velocity parameter $\mu_0$
for a linear trap superposed with a nonlinear barrier.
The theoretical trap condition is $\mu_T<0$
which gives $\mu_{crit} = 0.08$ for $V_{nl0}=1$, $V_{l0}=-1$ and
$\mu_{crit}=0.04$ for $V_{nl0}=0.5$,Ê$V_{l0}=-0.5$,
in good agreement with the simulations in the case $V_{nl0}=0.5$,Ê
$V_{l0}=-0.5$,.
\label{fig4} }
\end{figure}

By comparing the transmission through a linear trap in presence or
in absence of a nonlinear positive potential, we confirm the
numerical conjecture that the transmission coefficient can be
significantly increased by a nonlinear modulation (see Figures
\ref{fig3}-\ref{fig4}). In fact, the radiation emitted by the
soliton due to the interaction with the linear trap and with the
nonlinear potential can cancel each other, resulting in an
enhanced soliton transmittivity.

A similar result can be obtained with a linear barrier.
A nonlinear negative  modulation can help the soliton going through
the linear barrier by reducing the radiation emission and by
reducing the minimal velocity reached by the soliton
during the interaction. This was predicted by numerical simulations
in \cite{kevrekidis2}.

For consistency we propose an experimental configuration
where the  enhanced
transmission could be observed for the $^7$Li condensate.
The  transverse frequency
$\omega_{\perp} \approx 2\pi \times 10^3$~H and the density is $n =
10^9$~m$^{-1}$.
The healing length and speed of sound are $\xi \simeq 2$~$\mu$m
and $c \approx 5$~mm/s, respectively.
As a typical experiment, we
could consider a soliton with about $10^3$ atoms and width
$\approx 2\xi \approx 4$~$\mu$m
(so that  $\nu = 0.5$) and a linear trap with Gaussian shape,
normalized amplitude $0.5$ and width
$\approx \xi \approx 2$~$\mu$m (so that $x_c=0.5$).
The theoretical prediction is that such a soliton is trapped if its velocity
is below $0.55 c$ (because $\mu_{crit}\simeq 0.14$, see Figure \ref{fig3}).
However, in this experiment, we can consider variations of the
external magnetic field $B$ around the value $352$~G, where the
scattering length has the minimal value $\approx -0.23$~nm.
Increasing the field to the value $B = 450$G we can increase the
scattering length to the value $\approx -0.18$~nm.
This means that the scattering length can be varied by $25 \%$,
and thus a nonlinear barrier $V_{nl}$ with normalized amplitude $0.5$ can be generated.
The theoretical prediction is that, in the presence of the linear trap
and the nonlinear barrier, the soliton will be trapped
only if its velocity is below $0.2 c$
(because $\mu_{crit}\simeq 0.05$, see Figure \ref{fig4}),
and it will be transmitted otherwise.
This means that the nonlinear modulation dramatically enhances the
domain of parameters for which solitons can be transmitted through the linear trap.

\section{conclusion}
In this paper we have investigated the time-dependent nonlinear
scattering of bright solitonic matter waves through different
types of barriers. We have considered the transmission of matter waves
through inhomogeneities in the form of localized linear and
nonlinear potentials. The adiabatic dynamics of the wavepackets as
well as the radiative processes during the transmission have been
analyzed.

To analyze the dynamics we use  the perturbed IST
theory, which allows us to predict the trapping of a bright
soliton by a trap potential and the reflection of a soliton by a
barrier potential. The parameters (mass and velocity) of the
transmitted soliton can be estimated by computing the radiated
mass and energy and by using the conservations of the total mass
and energy.   The
enhanced transmission of a soliton through a linear trap by a
nonlinear modulation of the scattering length is explained by this
theory.

 The analytical predictions  have been checked by comparisons with
numerical simulations of the GP equation. The formulas are
valid in the asymptotic framework where the
potentials have small amplitudes. It turns out that, for small or
moderate potential amplitudes $|V_{l0}| \leq 0.5$, $|V_{nl0}| \leq
0.5$, the perturbed IST theory gives quantitatively accurate
predictions.
For large potential amplitudes $|V_{l0}| \geq 1$, $|V_{nl0}| \geq
1$, the theoretical predictions of the perturbed IST theory are
still in qualitative agreement with the numerical results. This
means that the perturbed IST theory is useful for probing the
parameter space and exhibiting interesting phenomena. The enhanced
transmission can be observed in the experiments with bright matter
wave solitons in elongated trap with proper variation in space of
the external magnetic field and the trap
potential \cite{kevrekidis,AG05}.
 One of the problems that should be addressed for
future consideration by this approach is the nonlinear resonant
scattering on the (periodic or random) chain of  nonlinear
barriers \cite{Sakaguchi} and the scattering on time-dependent
linear and nonlinear barriers \cite{Azbel}. The latter problem is
important for many areas of condensed matter.

\end{document}